\documentclass{elsart5p}

\usepackage{amsmath}
\usepackage{amssymb}
\usepackage{graphicx}
\usepackage{bm}
\usepackage{lineno}

\newcommand{\mrm}{\mathrm}
\newcommand{\sub}[1]{\mathrm{\scriptscriptstyle{#1}}}

\begin{document}

\begin{frontmatter}

\title{UCN production by multiphonon processes in\\ superfluid Helium under pressure}

\author[ILL,E18]{P. Schmidt-Wellenburg}
\ead{schmidt-w@ill.fr},
\author[ILL]{K.H.~Andersen},
\author[ILL,E18]{O. Zimmer}

\address[ILL]{Institut Laue Langevin, 6 rue Jules Horowitz,BP-156, 38042 Grenoble Cedex 9, France}
\address[E18]{ Physik-Department E18, Technische Universit\"{a}t M\"{u}nchen, 85748 Garching,
Germany}

\begin{abstract}
Cold neutrons are converted to ultra-cold neutrons (UCN) by the excitation of a single phonon or multiphonons in 
superfluid helium. The dynamic scattering function $ S(q,\hbar\omega)$ of the superfluid helium strongly depends 
on pressure, leading to a pressure-dependent differential UCN production rate. A phenomenological 
expression for the multiphonon part of the scattering function $s(\lambda)$ describing UCN production 
has been derived from inelastic neutron scattering data. When combined with the production rate from 
single phonon processes this allows us to calculate the UCN production for any incident neutron flux. For 
calculations of the UCN production from single phonon processes we propose to use the values 
$S^{\ast} = 0.118(8)$ at saturated vapour pressure and $S^{\ast} = 0.066(6)$ at 20~bar. As an example we will 
calculate the expected UCN production rate at the cold neutron beam for fundamental physics PF1b 
at the Institut Laue Langevin. We conclude that UCN production in superfluid helium under pressure 
is not attractive.
\end{abstract}


\end{frontmatter}

\section*{Introduction}

Current projects to increase the density of ultra-cold neutrons (UCN)
available for physics experiments employ neutron converters of superfluid
helium (He-II) \cite{Masuda2002,Baker2003,Zimmer2007} and solid deuterium 
\cite{Trinks2000,PSI2005,Saunders2004,Frei2007,Liu1998}. In these materials
cold neutrons with energies in the order of meV may be scattered down to the
neV energy range by the excitation of a single phonon or multiphonon in the
converter. The inverse process is suppressed by the Boltzmann factor if
the converter is kept at sufficiently low temperature. We consider UCN
production in He-II, for which first publications have focused on the single
phonon process, where incident $8.9$~\AA\ neutrons are scattered down to the
UCN energy range \cite{Golub1977,Abe2001}. Later two publications \cite%
{Korobkina2002,Schott2003} have set out to calculate the expected UCN
production rate from processes involving the emission of two or more
excitations but reached contradicting results.

The purpose of this article is to provide on the one hand a physical-model-based
extrapolation to short wavelengths of scattering data relevant for UCN
production. On the other hand we consider for the first time UCN production
in He-II under pressure. This is motivated by the pressure dependence of the
properties of He-II, in particular its dispersion curve and density.
Application of pressure to He-II increases the velocity of sound, such that
the dispersion curves of He-II and of the free neutron cross at shorter
neutron wavelength. For neutron beams from a neutron guide coupled to a
liquid deuterium cold source, the differential flux density $\mathrm{d}\Phi /%
\mathrm{d}\lambda $ in the range $8-9$~\AA\ normally increases for
decreasing wavelength. Pressure also increases the density of He-II. These
two facts may lead to the expectation that the single phonon UCN production rate increases with pressure. Furthermore, the multiphonon contributions might be
favourably affected by application of pressure. This justifies to investigate
whether pressure may provide an UCN source superior to a converter at
saturated vapour pressure (SVP).

The first part of this paper presents some general expressions for the UCN
production rate in He-II as given in an internal note in 1982 by Pendlebury 
\cite{Pendlebury/internal} and in Ref. \cite{Korobkina2002}. In the second
part we consider UCN production due to multiphonon scattering both for SVP
and for 20~bar, using inelastic scattering data for He-II measured at 0.5~K\,\cite{Gibbs1999}. We give a formula approximating the contribution from
multiphonon processes to the UCN production rate. This might be useful to
calculate, for any given spectrum of incident neutrons, the expected UCN
production rate. As an example, this is done here for the cold neutron beam
PF1b at the Institut Laue Langevin.

\section*{UCN production rate, general expressions}

UCN production in superfluid helium is due to coherent inelastic scattering
of the incident cold neutrons with energy $E$ (and wavenumber $k$) down to
energies $E^{\prime }$ ($k^{\prime }$). The maximum final energy is defined
by the wall potential of the converter vessel ($252(2)$~neV for beryllium)
with respect to the Fermi potential of He-II (18.5~neV at SVP):
$V_{\mathrm{c}} = 233(2)$~neV.
The production rate is given by

\begin{equation}
P_{\mathrm{UCN}}(V_{\mathrm{c}})=\int_{0}^{\infty }\mathrm{d}E\int_{0}^{V_{%
\mathrm{c}}}N\frac{\mathrm{d}\phi }{\mathrm{d}E}\cdot \frac{\mathrm{d}\sigma 
}{\mathrm{d}E^{\prime }}\left( E\rightarrow E^{\prime }\right) \,\mathrm{d}%
E^{\prime },  \label{eqn:production_1}
\end{equation}

\noindent where $\mathrm{d}\phi /\mathrm{d}E$ is the differential incident
flux, $N$ the helium number density,
and $\mathrm{d}%
\sigma /\mathrm{d}E^{\prime }$ the differential cross section for
inelastic neutron scattering. The latter is given by

\begin{equation}
\frac{\mathrm{d}\sigma }{\mathrm{d}E^{\prime }}=4\pi b^{2}\frac{k^{\prime }}{%
k}S(q,\hbar \omega ),  \label{eqn:ddcross_section}
\end{equation}

\noindent where $b$ is the neutron scattering length of $^{4}$He, $\hbar
\omega =E-E^{\prime }$, $q=k-k^{\prime }$, and $S(q,\hbar \omega )$ is the dynamic scattering
function evaluated for values on the
dispersion curve of the free neutron. Details of calculation can be found in
Ref.\,\cite{Korobkina2002} where the general result is given in
Eq.\,(9). We may write

\begin{equation}
P_{\mathrm{UCN}}(V_{\mathrm{c}})=N\,\sigma V_{\mathrm{c}}\frac{k_{\mathrm{c}}%
}{3\pi }\int_{0}^{\infty }\frac{\mathrm{d}\phi }{\mathrm{d}\lambda }%
\,s\left( \lambda \right) \lambda \,\mathrm{d}\lambda ,
\end{equation}

\noindent where $\sigma =1.34(2)$~barn \cite{Sears1992} is the scattering cross
section of $^{4}$He, $\hbar k_{\mathrm{c}}=\sqrt{2m_{\mathrm{n}}V_{\mathrm{c}%
}}$, and%

\begin{equation}
s(\lambda )=\hbar \int S(q,\hbar \omega )\delta (\hbar \omega -\hbar
^{2}k^{2}/2m_{\mathrm{n}})\mathrm{d}\omega  \label{s(lambda)}
\end{equation}

\noindent defines the UCN scattering function as function of the incident neutron
wavelength $\lambda $, making use of $q=k=2\pi /\lambda $, valid for $%
k^{\prime }\ll k$. We divide it into a single and a multiphonon part, $%
s(\lambda )=s_{\mathrm{I}}(\lambda )+s_{\mathrm{II}}(\lambda )$.

The single phonon contribution can be approximated by $s_{\mathrm{I}%
}(\lambda )=S^{\ast }\delta (\lambda ^{\ast }-\lambda )$, where $\lambda
^{\ast }=2\pi /q^{\ast }$ is the neutron wavelength at the intersection of
the dispersion curves of the free neutron and the helium ($q^{\ast }=0.706$~%
\AA$^{-1}$ for SVP), and $S^{\ast }=\hbar \int_{\mathrm{peak}}S(q^{\ast },\hbar
\omega )\mathrm{d}\omega $ denotes the intensity due to single phonon
emission. Evaluation of the single-phonon UCN production rate yields

\begin{equation}
P_{\mathrm{I}}(V_{\mathrm{c}})=N\sigma \left( \frac{V_{\mathrm{c}}}{E^{\ast }%
}\right) ^{3/2}\frac{\lambda ^{\ast }}{3}\beta \,S^{\ast }\left. \frac{%
\mathrm{d}\phi }{\mathrm{d}\lambda }\right\vert _{\lambda ^{\ast }},
\label{P_I}
\end{equation}

\noindent where $E^{\ast }=\hbar ^{2}q^{\ast 2}/2m_{\mathrm{n}}$, and the
Jacobian factor

\begin{equation*}
\beta =\frac{\left. \frac{\mathrm{d}E_{\mathrm{n}}}{\mathrm{d}q}\right\vert
_{q^{\ast }}}{\left. \frac{\mathrm{d}E_{\mathrm{n}}}{\mathrm{d}q}\right\vert
_{q^{\ast }}-\left. \frac{\mathrm{d}E_{\mathrm{He}}}{\mathrm{d}q}\right\vert
_{q^{\ast }}}
\end{equation*}

\noindent accounts for the overlap of the two dispersion curves, $E_{\mathrm{%
n}}(q)$ and $E_{\mathrm{He}}(q)$. For a beryllium-coated converter at SVP
this results in

\begin{equation}
P_{\mathrm{I}}(V_{\mathrm{c}})=4.97(38)\cdot 10^{-8}\frac{\text{\AA}}{\text{cm}}\left. \frac{\mathrm{d}%
\phi }{\mathrm{d}\lambda }\right\vert _{\lambda ^{\ast }}.
\end{equation}

\begin{table}[tbp]
\centering
\begin{tabular}{|l|r@{}l|r@{}l|}\hline
& \multicolumn{2}{c|}{SVP} &  \multicolumn{2}{c|}{0~bar} \\ \hline
$\lambda^{\ast}~[$\AA $]$ &    ~8&.92(2)~&~8&.26(2)~\\ 
$E^{\ast}$~[meV] &              1&.028 &   1&.20 \\ 
$N$~[$10^{22}\mathrm{cm^{-3}}$] & 2&.1835 &  2&.5317 \\ 
$\beta$ &                       1&.42(1) & 1&.21(1) \\ 
$S^{\ast}$ &                    0&.118(8) &0&.066(7) \\ 
\hline
relative factor &               1&       & 0&.41\\\hline
\end{tabular}%
\caption{Factors relevant for single phonon production rate deduced from
Ref.\thinspace \protect\cite{Abraham1970,Caupin2008,Gibbs1996}. An increase
in flux of a factor $\sim 2.5$, going from $8.9$~\AA \thinspace\ to $8.3$~\AA %
\thinspace\ is necessary to compensate for the loss in intensity.}
\label{tab:S1}
\end{table}

The differential multiphonon UCN production rate in the same units is given
by

\begin{equation}
\frac{\mathrm{d}P_{\mathrm{II}}\left( V_{\mathrm{c}}\right) }{\mathrm{d}%
\lambda }=N\sigma V_{\mathrm{c}}\frac{k_{\mathrm{c}}}{3\pi }\frac{\mathrm{d}%
\phi }{\mathrm{d}\lambda }\lambda s_{\mathrm{II}}(\lambda ),
\label{eq:multi_diff}
\end{equation}%
which depends explicitly on the form of $s_{\mathrm{II}}(\lambda )$ which is
discussed below.

\section*{UCN production rates from scattering data}

\begin{figure}[tbp]
\centering
\includegraphics[width=0.9\columnwidth]{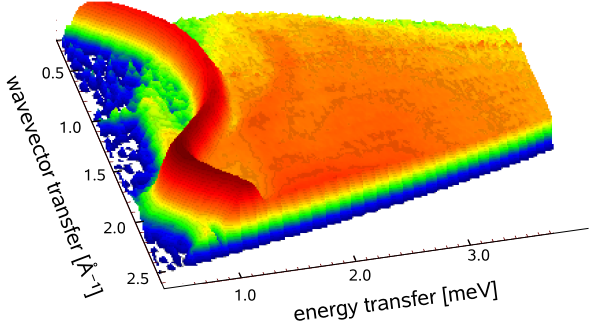}
\caption{Visualisation of $S(q,\hbar \protect\omega )$, for helium at $T=0.5$%
~K and $p=20$~bar, measured by\,\cite{Gibbs1999}.}
\label{fig:sqwplane}
\end{figure}

The scattering function Eq.\,(\ref{s(lambda)}) can be extracted
from inelastic scattering measurements. This has been done here for data
previously measured by Andersen and colleagues, on the IN6 time-of-flight
spectrometer at the Institut Laue Langevin, at
SVP and at $p\,=\,20$~bar. Parts of this data have already been published
earlier in \cite{Gibbs1999,Gibbs1996,Fak1991,Stirling1994,Gibbs2000}. For
the present purpose we have treated and rebinned all of the raw data to
obtain a higher resolution in $q$ and to cover the complete accessible $q$%
-range. Existing data points from previous publications were used as
reference for our analysis. A typical measurement of the dynamic scattering
function $S(q,\hbar \omega )$ is shown in Figure\,\ref{fig:sqwplane}%
.

\begin{figure}[tbp]
\centering
\includegraphics[width=\columnwidth]{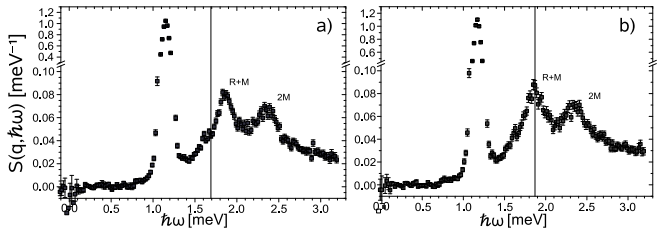}
\caption{$S(q,\hbar\protect\omega)$ for a)~$q = 0.90~\mathrm{\mathring{A}%
^{-1}}$ and b)~$q = 0.95~\mathrm{\mathring{A}^{-1}}$. The vertical lines
indicate the energy of an incident neutron with $E=\hbar^2 q^2/2 m_{\mathrm{%
n}}$ that can be down-scattered to the UCN energy
range. The width of the single phonon excitation is dominated by the finite
resolution of the instrument. The roton-maxon (R+M) and two maxon (2M)
resonances at higher energies are significantly lower in intensity.}
\label{fig:multiphonons}
\end{figure}

\noindent The plane is divided into slices with width $\Delta q=0.2~\mathrm{%
\mathring{A}^{-1}}$. Figure\,\ref{fig:multiphonons} shows $%
S(q,\hbar \omega )$ for two representative values of $q$. From each such slice we
extract one value of $s_{\mathrm{II}}(\lambda )$, with $\lambda $ fixed by
the condition $h/\lambda =\sqrt{2m_{\mathrm{n}}\hbar\omega}$. For this
purpose the energy is binned with width $\Delta \hbar \omega =0.02$~meV. The
compiled data is presented in Table\,\ref{tab:data_andersen} of the
appendix. For the short wavelength region below $4.5$~$\mathrm{\mathring{A}}$
only one data point exists from Ref.\,\cite{Fak1991}. An accepted model for the extrapolation
to short wavelengths is given by Family\thinspace \cite{Family1975}. It was
developed from low-order diagrams of the interparticle interactions in the
high frequency limit and predicts $S(q,\hbar \omega )\propto q^{4}\omega
^{-7/2}$, which leads to $s_{\mathrm{II}}(\lambda )\propto \lambda ^{3}$ in
our case.

\begin{figure}[tbp]
\centering
\includegraphics[width=\columnwidth]{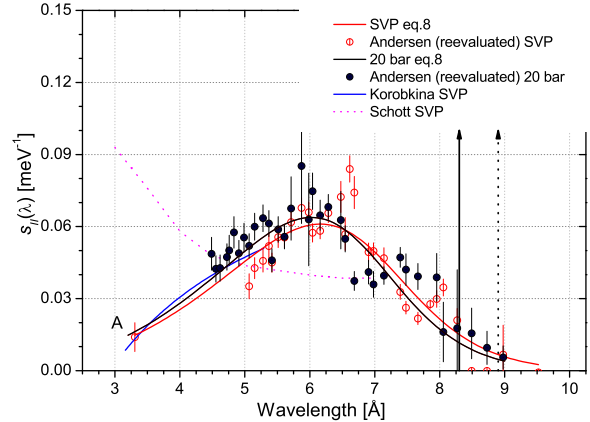}
\caption{Multiphonon scattering function at SVP and 20~bar. The extrapolation to short wavelength of Korobkina \textit{et al.}\,\cite{Korobkina2002} at SVP is linear in $k$, whereas the calculation of Schott \textit{et al.}\,\cite{Schott2003} is based on the static structure factor of superfluid helium. The data point (A) is taken from Ref.\,\protect\cite{Fak1991}. The one-phonon
peaks are indicated by vertical arrows: SVP~($\cdots$) and 20~bar~(---). The data for 20~bar include the corrections proposed in Ref.\,\cite{Caupin2008}.}
\label{fig:s2q}
\end{figure}

For a global characterisation of the multiphonon scattering function, we
have found the following analytical expression useful,

\begin{equation}
s_{\mathrm{II}}\left( \lambda \right) =\frac{f\lambda ^{3}}{\exp \left(
p\cdot \frac{E_{\mathrm{MR}}-h^{2}/2m_{\mathrm{n}}\lambda ^{2}}{E_{\mathrm{%
rec}}}\right) +1},  \label{eqn:model}
\end{equation}

\noindent where $E_{\mathrm{MR}}$ is the sum of the maxon and roton
energy. $E_{\mathrm{rec}}=\tfrac{\hbar^2q^2}{2m_{\sub{He}}}$ is the recoil energy of the He nucleus and $f$, $p$ are fitting
parameters which vary with pressure (see Table\,\ref{tab:fitting} for
values). For $\lambda <5$~\AA\, it converges to the model of
Family. In Figure\,\ref{fig:s2q} the correspondence of the model to
the measured data is shown for SVP and 20~bar.

Both the single and the multiphonon scattering functions change with
pressure. The single phonon excitation moves to shorter wavelength, $\beta $
and the value for $S^{\ast }$ decrease (see Eq.\,(\ref{P_I})).
Therefore the UCN production by single phonon processes decreases as shown
in Table\,\ref{tab:S1}, unless the cold neutron flux at $8.3$~\AA{} is a factor of $\sim 2.5$ greater than at $8.9$~\AA{} to compensate
for the loss in scattering intensity.

\begin{figure}[tbp]
\includegraphics[width=\columnwidth]{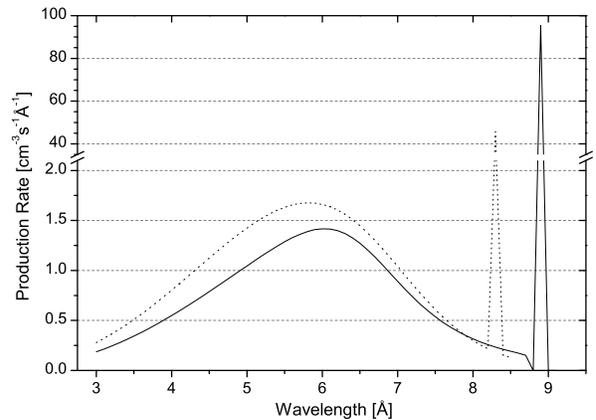}
\caption{Differential UCN production rate for SVP\,(---) and
20~bar\,(-\,-\,-) at PF1b. The calculations have been performed for a
critical wall potential of 252~neV. One can see clearly the shift
of the single phonon peak to shorter wavelength, as well as the subtle increase from multiphonon processes.}
\label{fig:prod}
\end{figure}

\noindent The situation looks slightly different for the UCN production rate
due to multiphonon scattering. With pressure the multiphonon excitations
increase in intensity and the broad peak centre moves to shorter
incident-neutron wavelengths (see Figure\,\ref{fig:s2q}). Integrating
over Eq.\,(\ref{eq:multi_diff}) with the incident cold neutron flux
of PF1b\thinspace \cite{PF1b2006} at the Institut Laue Langevin, one can
expect an increase in UCN production rate through multiphonon excitation.
The differential production rate is shown in Figure\,\ref{fig:prod}. Integration over the UCN production curves for SVP and 20~bar gives in
total $(13.9\pm 0.9)~\mathrm{cm^{-3}s^{-1}}$ and $(11.1\pm 0.8)~\mathrm{%
cm^{-3}s^{-1}}$, with a contribution of $(9.5\pm 0.7)~\mathrm{cm^{-3}s^{-1}}$
and $(5.8\pm 0.4)~\mathrm{cm^{-3}s^{-1}}$ from single phonon processes,
respectively.

\section*{Discussion}

We have shown how to calculate UCN production rates for different pressures
based on inelastic neutron scattering data. The single phonon UCN production
rate is slightly higher than in Ref.\thinspace \cite{Korobkina2002} as we
use for $S^{\ast}=0.118(8)$, instead of $S^{\ast}=0.1$. This higher value
from Gibbs \cite{Gibbs1996} (there called $Z(Q)$) is affirmed in a comparative
analysis in Ref.\thinspace \cite{Caupin2008}. A different approach by
Yoshiki \cite{Yoshiki2003}, calculating the total cross section for UCN
production from single phonon excitations in a beryllium coated converter,
gives a slightly larger value. However, for his numerical calculation he did
not take the Fermi potential of helium into account.

A phenomenological expression is given to describe the global behaviour of
the multiphonon scattering function. It converges for short incident
wavelength to the model of Family. This model describes the short wavelength
region better than the linear extrapolation in $q$ used in Ref.\thinspace 
\cite{Korobkina2002}, which overestimates the multiphonon contribution. The
calculations done by Schott and coworkers\thinspace \cite{Schott2003}
predict a huge increase in the scattering function for short wavelengths
that appears to contradict the first moment sum rule\thinspace \cite%
{Rahman1962}. Uncertainties remain in the region below $4$~\AA \thinspace\
as there is only one data point available\thinspace \cite{Fak1991}. Baker and coworkers\thinspace \cite{Baker2003} have measured steadily decreasing production
rates in the range of $6.5$~\AA{} to $4.5$~\AA{}. Therefore we assume to have only a small
contribution to the scattering function for $\leq 4$~\AA , which we consider
as being best represented by our expression. On the long wavelength end for $%
\lambda >\lambda ^{\ast }$ there is no intensity, as there are no more
elementary excitations in superfluid helium.

\section*{Conclusion}

UCN production in superfluid helium under pressure will not
lead to a gain in production rate compared to SVP using a cold neutron
spectrum similar to the one of the cold beam PF1b. Only if the flux at $8.3$~%
\AA \thinspace\ exceeds by more than 2.5 times that at $8.9$~\AA, an increase in UCN production rate is expected, due to the increase of
intensity of multiphonon processes. However, one has to bear in mind that
application of pressure requires a window for UCN extraction. Whereas, an
extraction window with associated severe UCN losses can be avoided for SVP
\cite{Masuda2002,Zimmer2007}.

\section*{Appendix}

\begin{table}[ht]
\centering
\begin{tabular}{|r@{.}l|r@{.}l@{\,$\pm$\,}r@{.}l|r@{.}l@{\,$\pm$\,}r@{.}l|}\hline

 \multicolumn{2}{|c|}{$q~[\mbox{\AA$^{-1}$}]$} & \multicolumn{4}{c|}{$s(q,20~\mathrm{bar})~[\mathrm{meV^{-1}}]$}& \multicolumn{4}{c|}{$s(q,\mathrm{SVP})~[\mathrm{meV^{-1}}]$}\\
\hline
~0&64 &\multicolumn{4}{c|}{}      &~0&00428 &0&00288~\\
0&66 &\multicolumn{4}{c|}{}      &0&1614  &0&00583\\
0&68 &\multicolumn{4}{c|}{}      &0&57394 &0&02023\\
0&70 &\multicolumn{4}{c|}{}      &0&9053  &0&01229\\
0&72 &~0&1000& 0&0069~&0&75145 &0&011\\
0&74 &0&2694& 0&0105&0&2743  &0&01227\\
0&76 &0&4633& 0&0282&0&06861 &0&00488\\
0&78 &0&3145& 0&0144&0&03653 &0&00351\\
0&79 &0&2455& 0&0115&0&03015 &0&00347\\
0&80 &\multicolumn{4}{c|}{}&0&02771 &0&00175\\
0&82 &0&0580& 0&0063&0&02175 &0&00255\\
0&84 &0&0420& 0&0076&0&02624 &0&00265\\
0&85 &0&0471& 0&0048&0&03282 &0&0031\\
0&88 &0&0395& 0&0042&0&04684 &0&00436\\
0&90 &0&0359& 0&0062&0&04984 &0&00338\\
0&91 &0&0410& 0&0057&0&04936 &0&00382\\
0&94 &0&0373& 0&0046&0&07415 &0&0067\\
0&95 & \multicolumn{4}{c|}{}&0&08398 &0&0055\\
0&96 &0&0548& 0&0063 &0&0569  &0&00684\\
0&97 &0&0626& 0&0138 &0&07238 &0&00648\\
1&00 &0&0681& 0&0059 &0&06567 &0&00394\\
1&02 &0&0646& 0&0073 &0&05824 &0&00336\\
1&04 &0&0746& 0&0088 &0&0574  &0&00384\\
1&05 &0&0628& 0&0223 &0&06596 &0&00425\\
1&07 &0&0851& 0&0277 &0&06774 &0&00419\\
1&10 &0&0674& 0&0153 &0&06188 &0&00442\\
1&12 &0&0556& 0&0058 &0&05543 &0&00417\\
1&14 &0&0587& 0&0061 &0&0556  &0&00431\\
1&16 &0&0459& 0&0055 &0&045   &0&004\\
1&17 &0&0612& 0&0062 &0&05191 &0&00555\\
1&19 &0&0634& 0&0064 &0&04576 &0&00472\\
1&22 &0&0599& 0&0065 &0&0427  &0&00439\\
1&24 &0&0519& 0&0058 &0&03515 &0&00518\\
1&26 &0&0553& 0&0068 &\multicolumn{4}{c|}{}\\
1&28 &0&0488& 0&0061 &\multicolumn{4}{c|}{} \\
1&30 &0&0575& 0&0076 &\multicolumn{4}{c|}{}\\
1&32 &0&0500& 0&0072 &\multicolumn{4}{c|}{}\\
1&33 &0&0473& 0&0064 &\multicolumn{4}{c|}{}\\
1&36 &0&0427& 0&0063 & \multicolumn{4}{c|}{}\\
1&38 &0&0423& 0&0063 &\multicolumn{4}{c|}{}\\
1&40 &0&0486& 0&0079 & \multicolumn{4}{c|}{}\\
1&90 &\multicolumn{4}{c|}{}      &0&014   &0&006\\\hline
\end{tabular} 
\caption{Scattering function $ s(q) = \hbar \int S(q,\hbar \omega) \delta(\hbar \omega - \hbar^2 q^2/2m_{\mrm{n}})\mrm{d}\omega$ evaluated for UCN production at SVP and 20~bar. Blank spaces indicate that no data points are available for these wave vectors.}
\label{tab:data_andersen}
\end{table}
        
\begin{table}[ht]
	\begin{tabular}{|l|r@{.}l@{}l@{.}l|r@{.}l@{}l@{.}l|}\hline
	    & \multicolumn{4}{c|}{SVP ($T=1.24$~K)} & \multicolumn{4}{c|}{20~bar~($T=0.5$~K)} \\
	\hline 
	  $f$ [$10^{-3}$\AA$^{-3}$meV$^{-1}$] &~0&3 &\,$\pm$\,0&01&  ~0&6&\,$\pm$\,0&02\\  
	  $p$                                 & 4&46&\,$\pm$\,0&26 & 1&12&\,$\pm$\,0&09\\
	$E_{\sub{MR}}$~[$\mrm{meV^{-1}}$]& 1&93&\multicolumn{2}{c|}{}&1&94&\multicolumn{2}{c|}{} \\\hline
	\end{tabular}
	\caption{$E_{\sub{MR}}$ from Ref.\,\cite{Gibbs1996} and fitting parameters of Eq.\,(\ref{eqn:model}) for SVP and 20~bar, determined in a least--$\chi^2$ fit to the inelastic scattering data.}
	\label{tab:fitting}
        
\end{table}

\end{document}